# RESEARCH ON THE SECONDARY ELECTRON YIELD OF TiZrV-Pd THIN FILM COATINGS

WANG Jie（王洁）, XU Yan-Hui（徐延辉）, ZHANG Bo（张波）, WANG Yong（王勇）[1)]  WEI Wei（尉伟）

National Synchrotron Radiation Laboratory，University of Science and Technology of China, HeFei, AnHui 230029 China

\* Work supported by the National Nature Science Foundation of China under Grant Nos. 11475166.

1） corresponding author: ywang@ustc.edu.cn

## ABSTRACT

In particle accelerators, the build-up of electron cloud may have important influence on beam quality. Especially for the positron and proton accelerators, massive electrons lead to electron cloud, which affects the stability, energy, emittance and beam life adversely. A secondary electron emission (SEE) measurement system has been designed and used to study the SEE of palladium (Pd), TiZrV and TiZrV-Pd with an independently adjustable energy from 50 eV to 5 keV. Here, we obtained the characteristics of the SEE from Pd, TiZrV and TiZrV-Pd film coatings with different thickness under ultrahigh-vacuum (UHV) conditions. Moreover, the maximum secondary electron yield (SEY), $\delta_{max}$, of the Pd, TiZrV and TiZrV-Pd film coatings under different primary electron doses were obtained, respectively. Finally, the variation of the secondary electron yield with the incident electron energy will be discussed for Pd, TiZrV and TiZrV-Pd thin film coatings. Low SEY is a new advantage of TiZrV-Pd films, besides high $H_2$ absorption ability and prolonging the lifetime of TiZrV film, which will be of great value in the design of beam screen for Super Proton-Proton Collider (SPPC).

Key words: secondary electron yield (SEY), TiZrV-Pd, low SEY film coating, secondary electron emission (SEE)
PACS: 29.20.-c Accelerators

## 1 INTRODUCTION

Emission from the vacuum chamber after impact from incident particles or synchrotron radiation or ionization of residual gas particles in the vacuum chamber, will induce the increase of the number of free electrons in an accelerator vacuum chamber and form the electron cloud. The build-up of electron cloud in the beam pipes may considerably hinder the stability of the high-intensity particle beams. In particular, increasing the intensity of the beam means that better suppression of secondary electron emission is needed in the beam pipes. The test of secondary Electron Yield (SEY) of the vacuum chamber material is important and SEY has great influence on the process of free electrons building up. Over the past few decades, many research institutions, such as CESR [1], CERN [2], SLAC [3], FERMILAB [4], KEKB[5] have done some research on secondary electron emission measurement.

A series of in-situ measurements data of the SEY have been obtained at Cornell [6-8] and FERMILAB [9]. The SEY of materials which are located at the beam pipe's wall can be measured in the environment of a running accelerator. The in-situ SEY test system is also for periodic measurements to observe beam conditioning of the SEY and discrimination between exposure to direct photons from synchrotron radiation versus scattered photons and cloud electrons [8]. However, in-situ SEY test system is expensive and need to be measured under the condition of a running accelerators.

Other SEY test set-up is independent and widely used in many labs, such as KEKB [10], CERN [11]. It is easy to measure the SEY and does not restricted by the accelerator. The SEY of materials is influenced by the material property, surface condition, topography, electron energy and electron dose etc., so separate SEY test set-up usually combine with Auger electron spectroscopy (AES) and static-secondary-ion mass spectroscopy (SSIMS) to monitor the corresponding variations in the surface chemical composition.

Here we designed and used separate SEY test set-up to compare TiZrV and TiZrV-Pd films with different thickness deposited by direct current (DC) magnetron sputtering. Pd thin film coatings are added onto the TiZrV film that is mainly applied in the ultra-high vacuum (UHV) pipes of storage rings to increase the service life of non-evaporable getters and enhance $H_2$ pumping speed. Several researchers have studied the absorbing behavior and preparation procedure of TiZrV-Pd, such as Mura [12, 13], Benvenuti [14] etc. Nonetheless, SEY data for TiZrV-Pd film coatings are rare and the effect of thickness, substrate, electrons energy and electrons dose on the SEY variation is deserving research.

In this article, the TiZrV-Pd films were characterized for microstructural and surface roughness, and finally the SEY for the films with different thickness was investigated. The SEY test-stand, is capable of measuring the SEY from samples using an incident electron beam when the samples are biased at different voltages.

## 2 Experimental

### *2.1 Experimental set-up and procedures*

Kimball Physics EGL-2022 electron gun were installed and directed towards the sample at a 90°. The electron gun scans over an energy spectrum of 50 eV to 5000 eV on the samples. A Keithley 2400 pico-Ammeter which have an accuracy of 0.012%, is used to indirectly measure the SEY of the sample and apply bias voltage during measurements. So, for all SEY values mentioned in this paper, the error of SEY is 0.024%. The vacuum vessel is grounded during measurements. The emission current of the gun was set as 1μA, 2μA, 5μA, 10μA respectively at Emission Current Control (ECC) mode. These samples were delivered under nitrogen and immediately installed in the vacuum chamber after opening. All measurements were performed at $10^{-9}$ torr. The temperature of the samples during the tests is about 300 K.

The SEY is the ratio of the number of secondary electrons emitted from a surface, $I_{\text{SEY}}$, to the number of electrons incident to that surface, $I_P$, which is measured by applying a +150 V bias voltage that recaptures all secondary electrons. The total current $I_t$ is measured by applying -40 V bias voltage that repels all low energy secondary electrons. So the secondary emission current is given by $\boldsymbol{I}_{\text{SEY}} = \boldsymbol{I}_t - \boldsymbol{I}_P$. Therefore, the SEY can be calculated by the following equation. The SEY test device is shown in Fig. 1.

$$I_{\text{SEY}} = \frac{\boldsymbol{I}_{\text{SEY}}}{\boldsymbol{I}_p} = 1 - \frac{I_t}{I_p}$$

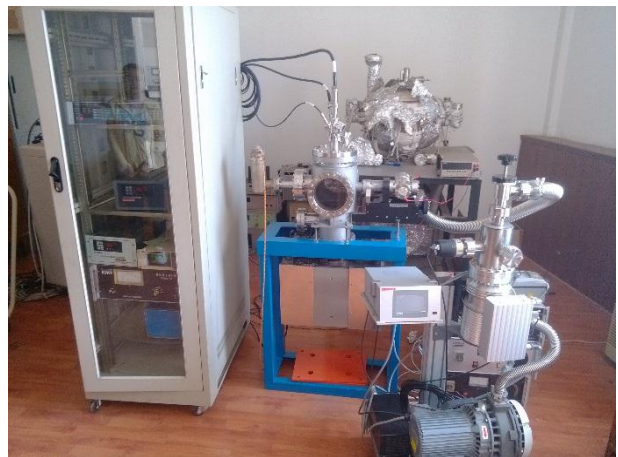

Figure 1: SEY test device.

## 2.2 Sample preparation

The TiZrV, TiZrV-Pd, Pd samples tested were in the form of thin films, which were used as film coatings in ultra-high vacuum pipes for particle acceleration, shown in Table 1. Different thickness of Pd thin films was deposited onto a polished Silicon substrate by magnetron sputtering inside a stainless steel pipe. After sputtering deposition, the TiZrV-Pd thin films were vented to atmospheric pressure with dry $N_2$. The samples were exposure to laboratory air before the sample introduction into the vacuum chamber of the SEY test systems. The thickness and section morphology of the TiZrV-Pd, TiZrV, Pd thin films were observed with a scanning electron microscope (SEM). And, the surface roughness of these films were tested by Atomic Force Microscope (AFM).

Table 1: The thickness and surface roughness of TiZrV-Pd, TiZrV and Pd films samples.

| Sample | #TiZrV-Pd-A | #TiZrV-Pd-B | #TiZrV-Pd-C | #TiZrV-A | #TiZrV-B | #TiZrV-C | #Pd-A | #Pd-B | #Pd-C |
|---|---|---|---|---|---|---|---|---|---|
| Thickness /nm | 148 | 200 | 370 | 375 | 814 | 1470 | 150 | 248 | 438 |
| Roughness /nm | 6.7 | 44.4 | 17.4 | 4.6 | 6.2 | 13.9 | 7.9* | 10.5 | 9.9 |

## 3 Results and discussion

The SEY results as a function of energy of the primary electrons are shown in Fig. 2~4 for samples of TiZrV-Pd, TiZrV and Pd films, respectively, with different emission current and thickness. These dependences can be described in terms of a maximum value of SEY, $\delta_{max}$, measured at corresponding primary electron energy $E_{max}$ and the incident charge per unit area(Q), shown in Table 2. In addition, the primary current on the surface of the samples is corresponding to the Q, shown in the last column of Table 2. It can be seen that the maximum $\delta_{max}$ of the TiZrV-Pd, TiZrV and Pd samples are 1.82, 1.96, 1.83, and the minimum $\delta_{max}$ of them are 1.38, 1.51, and 1.52, respectively.

Table 2: The $\delta_{max}$ and $E_{max}$ of TiZrV-Pd, TiZrV and Pd films samples under different primary electron doses.

| Sample | $\delta_{max}$ | $E_{max}$ /eV | Q /C·mm$^{-2}$ | Primary current/μA |
|---|---|---|---|---|
| #TiZrV-Pd-A | 1.61 | 420 | $1*10^{-5}$ | 1 |
| | 1.54 | 600 | $2*10^{-5}$ | 2 |
| | 1.47 | 600 | $5*10^{-5}$ | 5 |
| | 1.38 | 650 | $1*10^{-4}$ | 10 |
| #TiZrV-Pd-B | 1.69 | 650 | $1*10^{-5}$ | 1 |
| | 1.75 | 600 | $2*10^{-5}$ | 2 |
| | 1.62 | 530 | $5*10^{-5}$ | 5 |
| | 1.52 | 700 | $1*10^{-4}$ | 10 |
| #TiZrV-Pd-C | 1.82 | 400 | $1*10^{-5}$ | 1 |
| | 1.81 | 600 | $2*10^{-5}$ | 2 |
| | 1.65 | 500 | $5*10^{-5}$ | 5 |
| | 1.55 | 570 | $1*10^{-4}$ | 10 |
| #TiZrV-A | 1.81 | 320 | $1*10^{-5}$ | 1 |
| | 1.71 | 470 | $2*10^{-5}$ | 2 |
| | 1.58 | 420 | $5*10^{-5}$ | 5 |
| | 1.68 | 200 | $1*10^{-4}$ | 10 |
| #TiZrV-B | 1.87 | 320 | $1*10^{-5}$ | 1 |
| | 1.56 | 400 | $2*10^{-5}$ | 2 |

| | 1.54 | 450 | $5*10^{-5}$ | 5 |
| --- | --- | --- | --- | --- |
| | 1.54 | 370 | $1*10^{-4}$ | 10 |
| #TiZrV-C | 1.96 | 350 | $1*10^{-5}$ | 1 |
| | 1.72 | 470 | $2*10^{-5}$ | 2 |
| | 1.57 | 470 | $5*10^{-5}$ | 5 |
| | 1.51 | 470 | $1*10^{-4}$ | 10 |
| Pd-A | 1.77 | 450 | $1*10^{-5}$ | 1 |
| | 1.72 | 570 | $2*10^{-5}$ | 2 |
| | 1.62 | 570 | $5*10^{-5}$ | 5 |
| | 1.52 | 650 | $1*10^{-4}$ | 10 |
| Pd-B | 1.82 | 420 | $1*10^{-5}$ | 1 |
| | 1.83 | 570 | $2*10^{-5}$ | 2 |
| | 1.71 | 600 | $5*10^{-5}$ | 5 |
| | 1.63 | 570 | $1*10^{-4}$ | 10 |
| Pd-C | 1.68 | 650 | $1*10^{-5}$ | 1 |
| | 1.72 | 530 | $2*10^{-5}$ | 2 |
| | 1.59 | 570 | $5*10^{-5}$ | 5 |
| | 1.62 | 570 | $1*10^{-4}$ | 10 |

## *3.1 TiZrV-Pd films*

Fig. 2 depicts that for sample #TiZrV-Pd-A with a thickness of 148 nm, $\delta_{max}$ decreased from 1.61 to 1.38 when the Q increased from $1\times10^{-5}$ to $1*10^{-4}$ C•mm$^{-2}$. Furthermore, it shows the same trend for sample #TiZrV-Pd-C with a thickness of 370 nm. However, the maximum $\delta_{max}$ of sample #TiZrV-Pd-C with a thickness of 200 nm was 1.75 when Q was $2*10^{-5}$ C•mm$^{-2}$, and then the maximum $\delta_{max}$ decreased to 1.52 when Q increased from $2*10^{-5}$ to $1*10^{-4}$ C•mm$^{-2}$. Therefore, for TiZrV-Pd film coatings with different thickness, $\delta_{max}$ does not always decreased when Q increased from $1*10^{-5}$ to $1*10^{-4}$ C•mm$^{-2}$. For the same Q, the $\delta_{max}$ of 148nm-#TiZrV-Pd-A film with a roughness of 6.7 nm is the lowest and 370 nm-#TiZrV-Pd-C with a roughness of 17.4 is the highest in these three samples, #TiZrV-Pd-A, #TiZrV-Pd-B and #TiZrV-Pd-C.

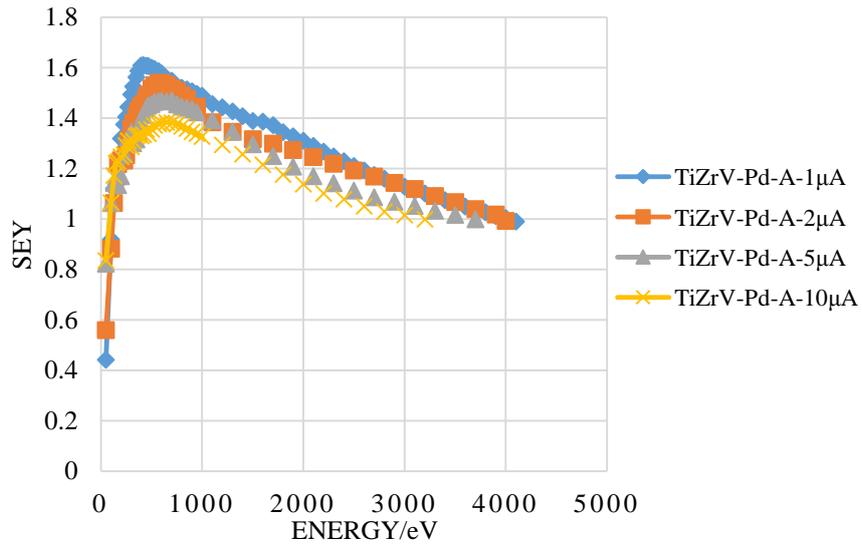

(a) #TiZrV-Pd-A

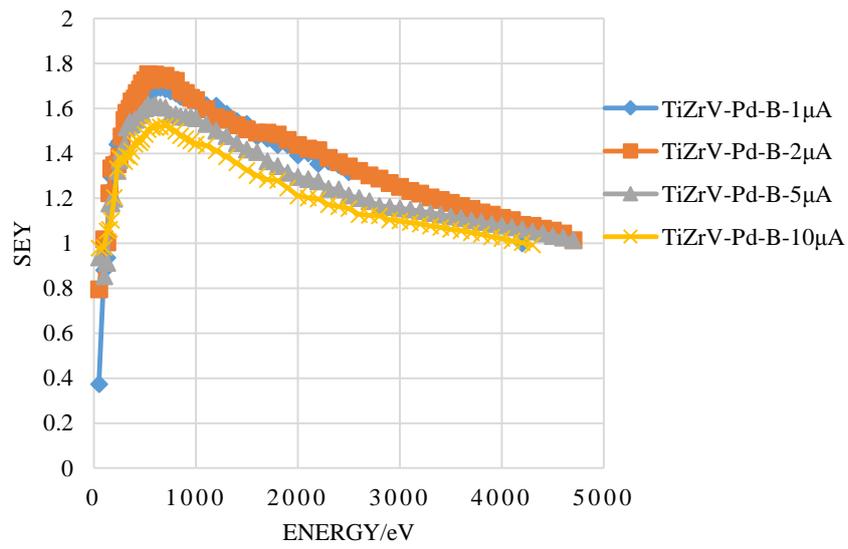

(b) #TiZrV-Pd-B

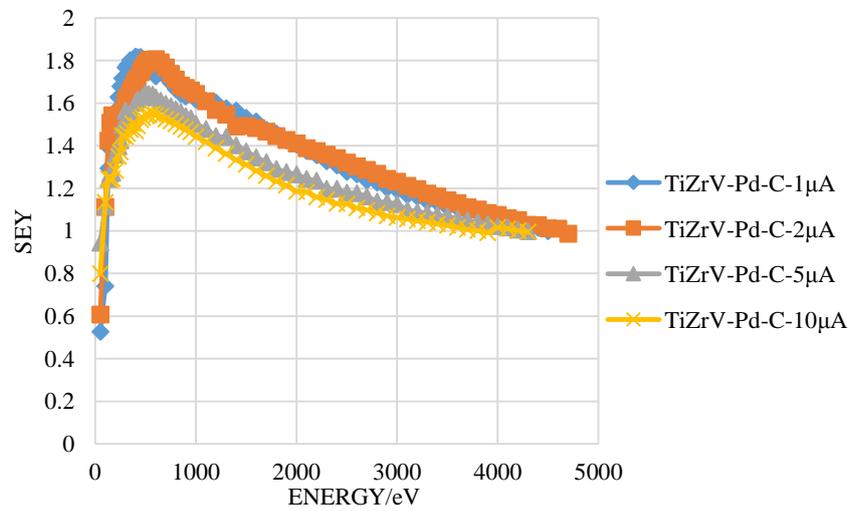

(c) #TiZrV-Pd-C

(c) #TiZrV-Pd-C

Figure 2: SEY for TiZrV-Pd film coatings as a function of incident electron energy: TiZrV-Pd-A —148 nm, TiZrV-Pd-B —200 nm, TiZrV-Pd-C —370 nm, and conditioning—electron bombardment with a dose of $1*10^{-5}$, $2*10^{-5}$, $5*10^{-5}$, $1*10^{-4}$ C•mm$^{-2}$.

*3.2 TiZrV films*

Fig. 3 states that for sample #TiZrV-A with a thickness of 375 nm, $\delta_{max}$ decreased from 1.81 to 1.58 when Q increased from $1*10^{-5}$ to $5*10^{-5}$ C•mm$^{-2}$ and then increased when Q increased to $1*10^{-4}$ C•mm$^{-2}$. For sample #TiZrV-C with a thickness of 1470 nm, $\delta_{max}$ decreased from 1.96 to 1.51 when Q increased from $1*10^{-5}$ to $1*10^{-4}$ C•mm$^{-2}$. However, for sample #TiZrV-B with a thickness of 814 nm, the difference is that $\delta_{max}$ were the same, 1.54, when Q increased from $5*10^{-5}$ to $1*10^{-4}$ C•mm$^{-2}$. The roughness of #TiZrV-A, #TiZrV-B, and #TiZrV-C thin films are 4.6 nm, 6.2 nm, 13.9 nm, respectively, with a scanning range of 5$\mu$m. The error of the surface roughness of #TiZrV-A and #TiZrV-B samples is 1.6 nm, so the effect of surface roughness on SEY can be ignore. Furthermore, the thickness of #TiZrV-B is two times of #TiZrV-A. When the values of Q are $2*10^{-5}$, $5*10^{-5}$ and $1*10^{-4}$, respectively, the $\delta_{max}$ of #TiZrV-B is lower than that of #TiZrV-A. However, the exception is that when Q is $1*10^{-5}$ C•mm$^{-2}$, the $\delta_{max}$ of TiZrV films increase from 1.81 to 1.96 with the increase of film thickness.

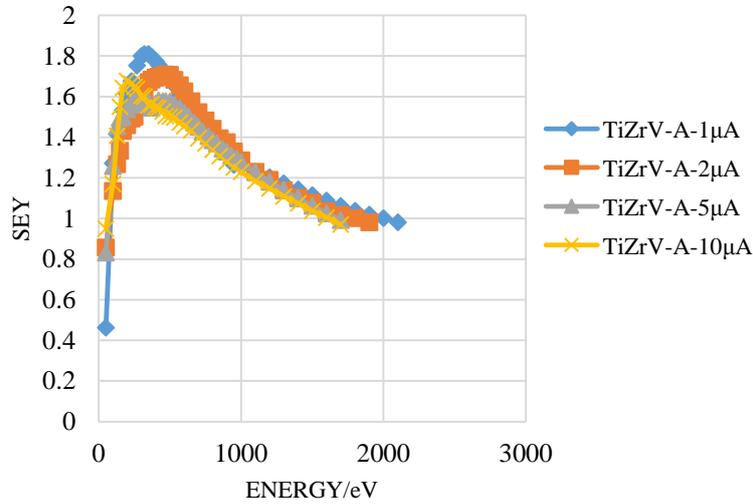

(a) #TiZrV-A

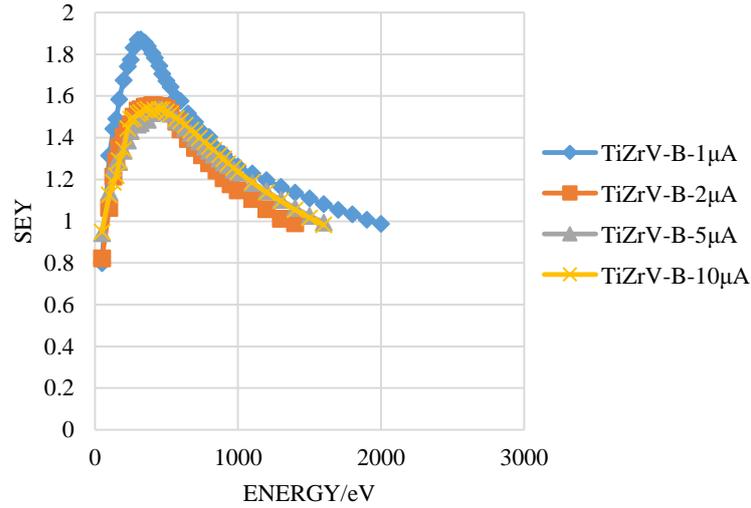

(b) #TiZrV-B

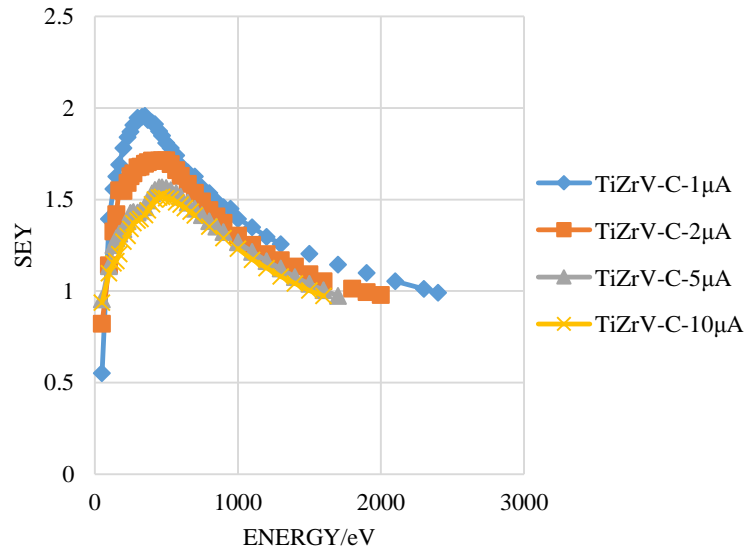

(c) #TiZrV-C

Figure 3: SEY for TiZrV film coatings as a function of incident electron energy: #TiZrV-A —375 nm, #TiZrV-B —814 nm, #TiZrV-C —1470 nm, and conditioning—electron bombardment with a dose of $1*10^{-5}$, $2*10^{-5}$, $5*10^{-5}$, $1*10^{-4}$ C•$mm^{-2}$.

## *3.3 Pd films*

Fig. 4 illustrates that for sample #Pd-A with a thickness of 150 nm, $\delta_{max}$ decreased from 1.77 to 1.52 when Q increased from $1*10^{-5}$ to $1*10^{-4}$ C•$mm^{-2}$. What is more, it shows the same trend for sample #Pd-B with a thickness of 248 nm. But when Q increased from $1*10^{-5}$ to $2*10^{-5}$ C•$mm^{-2}$, $\delta_{max}$ increased from 1.82 to 1.83. The roughness of # Pd-A, # Pd-B, and # Pd-C thin films are 7.9 nm, 10.5 nm, 9.9 nm, respectively, with a scanning range of 5$\mu$m, so the influence of Pd film roughness on $\delta_{max}$, by and large, can ignore.

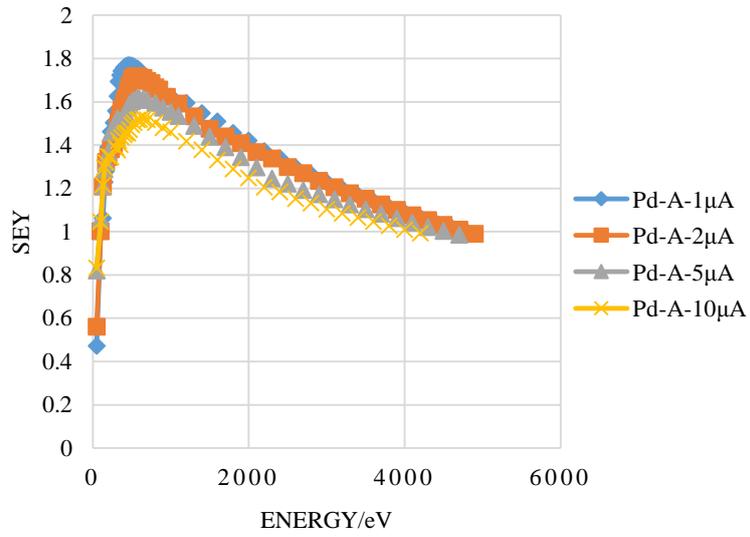

(a) #Pd-A

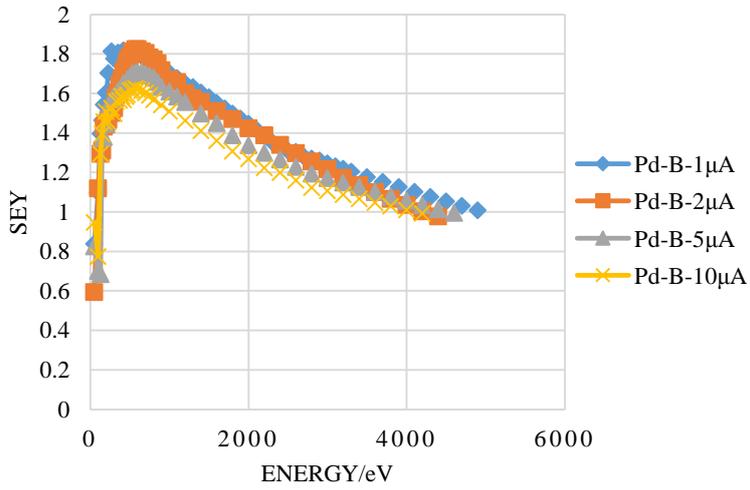

(b) #Pd-B

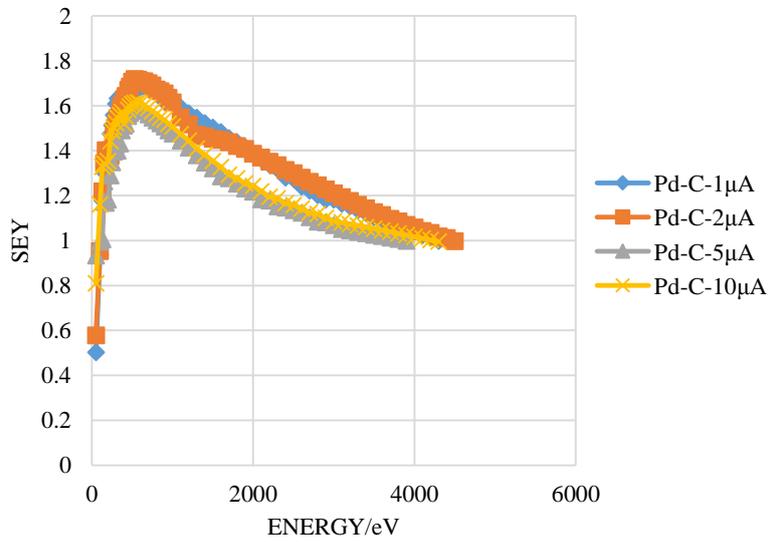

(c) #Pd-C

Figure 4: SEY for Pd film coatings as a function of incident electron energy: #Pd-A —150 nm, #Pd-B —248 nm, #Pd-C —438 nm, and conditioning—electron bombardment with a dose of $1*10^{-5}$, $2*10^{-5}$, $5*10^{-5}$, $1*10^{-4}$ C•mm$^{-2}$.

The $\delta_{max}$ of the Pd films with TiZrV substrates, are between 1.38 and 1.82, and they are 1.52~1.83 for Pd films with silicon substrates, shown in Fig. 2 and Fig. 4. The thickness of sample #TiZrV-Pd-A and #Pd-A are 148 and 150 nm, respectively, basically the same. When Q is $1*10^{-5}$ C•mm$^{-2}$, the $\delta_{max}$ of sample #TiZrV-Pd-A is lower than that of sample #Pd-A. The rule is the same, when Q is $2*10^{-5}$ C•mm$^{-2}$, $5*10^{-5}$ C•mm$^{-2}$ and $1*10^{-4}$ C•mm$^{-2}$. Therefore, from the experimental data it can be excluded that the maximum SEY of Pd films with TiZrV substrate is lower than that with silicon substrate, when the thickness of Pd films are the same.

According to the experimental results, the lowest $\delta_{max}$ of TiZrV films is 1.51, and it is 1.38 for Pd films, shown in table 2. So, lower SEY is a new advantage of TiZrV-Pd films, besides high $H_2$ absorption ability and prolonging the lifetime of TiZrV film. This result means that TiZrV-Pd film would be a promising material for the construction of beam screen for SPPC [15].

In order to realize the transport of primary electrons and the generation of secondary electrons, CASINO software [16] was used, shown in Fig. 5. Secondary electrons are mainly created where most of primary electrons lose their energy. When the energy of primary electrons varies between 100 eV and 4000 eV, the penetration depth $Z_{max}$ of primary electrons varies from 3 nm to 82 nm. According to reference [17], $\delta$ has a maximum value $\delta_{max}$ at a certain primary energy $E_{max}$ with a corresponding range near $Z_{max} \approx \lambda$, where $\lambda$ is the mean electron escape depth. Therefore, when $Z_{max}$ is about 8 nm at 600 eV, $\delta$ of Pd film has a maximum value. At high energies most secondary electrons are generated at a depth greater than λ and therefore do not escape the surface, leading to the decreasing of δ with increasing incident energy.

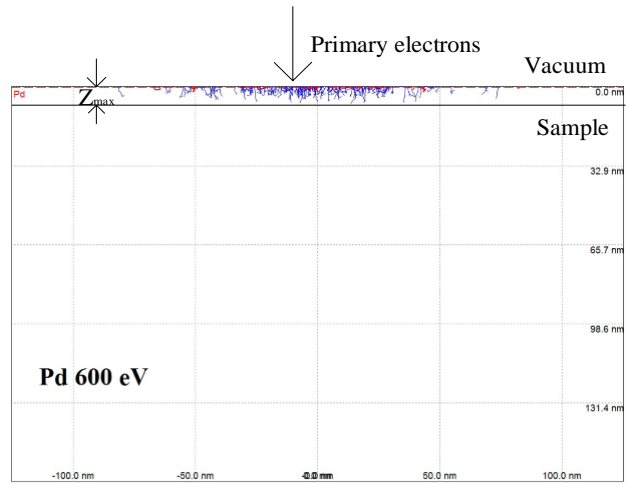

(a)

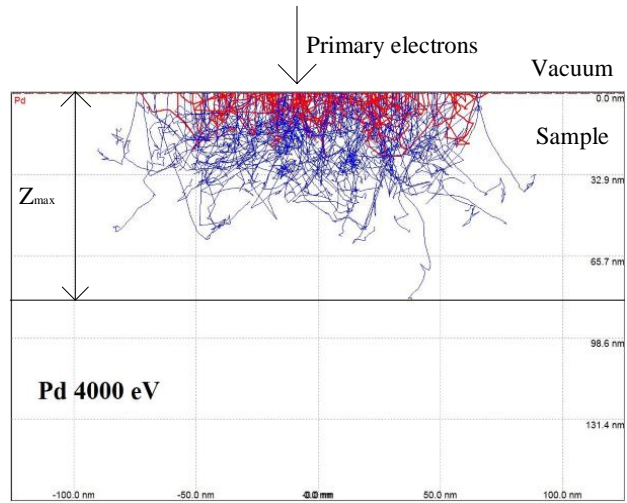

(b)

Figure 5: Primary electron trajectories in the Pd film with a thickness of 148 nm. Monte Carlo simulation results obtained with CASINO (the red trajectories represent backscattered electrons, the blue trajectories represent secondary electrons).

## 4 CONCLUSION

In conclusion, it is noticeable that the TiZrV-Pd, TiZrV, Pd thin films, have different SEY-versus-energy characteristics. The minimum $\delta_{max}$ of TiZrV-Pd, TiZrV, Pd thin films are 1.38, 1.51, and 1.52 and the thickness of the related films are 148 nm, 1470 nm, 150 nm, respectively. It is important to choose the appropriate thickness of TiZrV and TiZrV-Pd film coatings for e-cloud mitigations. Moreover, lower SEY would be a new advantage for TiZrV-Pd films, besides high $H_2$ absorption ability and prolonging the lifetime of TiZrV film. The results shown in this article will be of great value for beam screen construction of next generation accelerators, such as SPPC.